\documentclass{iopconfser}

\usepackage{amssymb}
\usepackage{ graphicx}

\begin{document}

\title{Review on Axions}

\author{Andreas Ringwald$^{1}$}

\affil{$^1$Deutsches Elektronen-Synchrotron DESY, Notkestr. 85, 22607 Hamburg, Germany}

\email{andreas.ringwald@desy.de}

\begin{abstract}
This proceedings' contribution explores the rationale behind the axion as a resolution to the strong CP puzzle. 
It outlines various benchmark axion models and examines their implications, focusing on two key aspects: 
(i) the axion's interactions with the Standard Model particles, and (ii) its potential role as dark matter. 
Additionally, it provides an overview of the discovery prospects associated with ongoing and future axion experiments. 
These efforts span from laboratory-based endeavors aimed at directly producing and detecting axions to searches for solar axions, 
and finally, to the direct detection of axion dark matter.
\end{abstract}

\section{The Axion}

\subsection{The strong CP puzzle}

The axion emerges within extensions of the Standard Model (SM) that elegantly address two fundamental questions simultaneously: i) the enigma of dark matter (DM), and ii) the remarkable precision of time reversal ($T$) and parity ($P$) invariance in strong interactions. 
Notably, experimental measurements of the most sensitive observable of $T$ and $P$ violation in flavor-conserving interactions -- the electric dipole moment  of the neutron (nEDM) -- have thus far only yielded an upper bound: $|d_n| < 1.8\times 10^{-26}\,e\,{\rm cm}$~\cite{Abel:2020pzs}. 
In contrast, from dimensional analysis, one expects an nEDM of order\footnote{Here, $e$ is the unit of electric charge, and $m_n$ is the mass of the neutron.}  $d_n \sim e/m_n \sim 10^{-13}\,e\,{\rm cm}$, if strong interactions violate $T$ and $P$.
Intriguingly, soon after the proposition of Quantum Chromo-Dynamics (QCD) as the fundamental theory of strong interactions, it was revealed that its most general Lagrangian encompasses a term which violates both $T$ and $P$, and thus $CP$, where $C$ denotes charge conjugation: the $\theta$-term, 
\begin{equation}
{\mathcal L}_{\rm QCD}\supset 
\overline\theta\,\frac{\alpha_s}{8\pi}   
  G_{\mu\nu}^b \tilde{G}^{b,\mu\nu}\,.
\label{eq:theta_term}
\end{equation}
Here, $\alpha_s=g_s^2/(4\pi)$ is the strong fine-structure constant in terms of the strong coupling $g_s$, $G_{\mu\nu}^b$ ($\tilde{G}^{b,\mu\nu}$) is
the (dual) gluonic field strength tensor, and $\overline\theta\in [-\pi,\pi]$ is an angular parameter, which serves as a measure of 
the strength of $T$ and $P$ violation in QCD. 
It induces an nEDM~\cite{Crewther:1979pi} approximately of the order $d_n \sim \bar\theta\,e\,m_*/m_n^2  \sim 10^{-16}\ \bar\theta\ e\,{\rm cm}$,  where $m_*=m_u m_d/(m_u+m_d)$ denotes the reduced mass of the $u$ and $d$ quarks. Consequently, the experimental upper bound imposes a stringent constraint on the  $\overline\theta$-parameter: 
$|\overline\theta| \lesssim 10^{-10}$.  The unnatural small value of this quantitiy constitutes the strong $CP$ puzzle.

\subsection{Minimal SM extension solving the strong CP puzzle}

Kim~\cite{Kim:1979if}, Shifman, Vainshtein, and Zakharov~\cite{Shifman:1979if}  (KSVZ) proposed a minimal renormalizable extension of the SM that effectively tackles the strong $CP$ puzzle. 
This extension introduces a new complex scalar field $\sigma$, which is a singlet under the SM, featuring a global axial 
Peccei-Quinn~\cite{Peccei:1977hh} (PQ) $U(1)$ symmetry. 
This symmetry is assumed to be spontaneously broken at a scale $v_{\rm PQ}$ much higher than the electroweak scale, 
as ensured by a postulated Mexican hat like potential term  in the Lagrangian, 
$$\mathcal{L}_{\rm KSVZ} \; \supset \; - \lambda_{\sigma} \left( \left| \sigma \right|^{2} -{v_{\rm PQ}^2}/{2} \right)^{\!\! 2}.$$
Additionally, a novel quark, denoted as $\mathcal Q$, is introduced, which carries charge under the $U(1)_{\rm PQ}$ symmetry and interacts with 
the complex PQ scalar $\sigma$ via a Yukawa coupling: 
$$\mathcal{L}_{\rm KSVZ} \; \supset \;   y \, \sigma \, \bar{\mathcal Q}_L {\mathcal Q}_R + {\rm h.c.}.$$ 
Utilizing a parameterization of the complex field $\sigma$ in terms of its modulus and phase, expressed as $\sigma (x) =  (1/{\sqrt{2}})\left( v_{\rm PQ} + \rho (x)\right) {\rm e}^{i a(x)/v_{\rm PQ}}$, we can discern that this model introduces three particles beyond the SM framework. Among them, two, denoted as $\rho$ and $\mathcal  Q$ , exhibit significant mass, approximately on the scale of the PQ breaking scale. 
%
\begin{figure}[h]
\centerline{\includegraphics[width=0.33\linewidth]{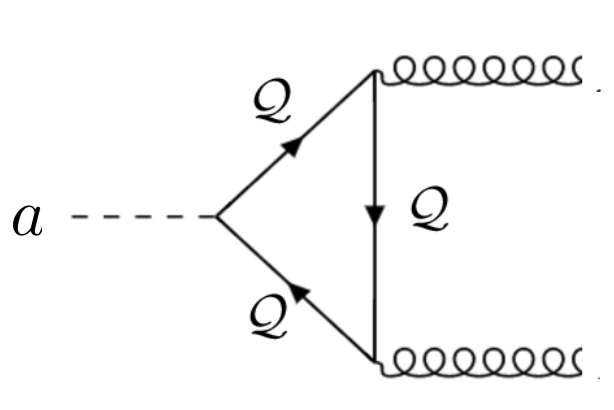}
\hspace{10ex}
                  \includegraphics[width=0.22\linewidth]{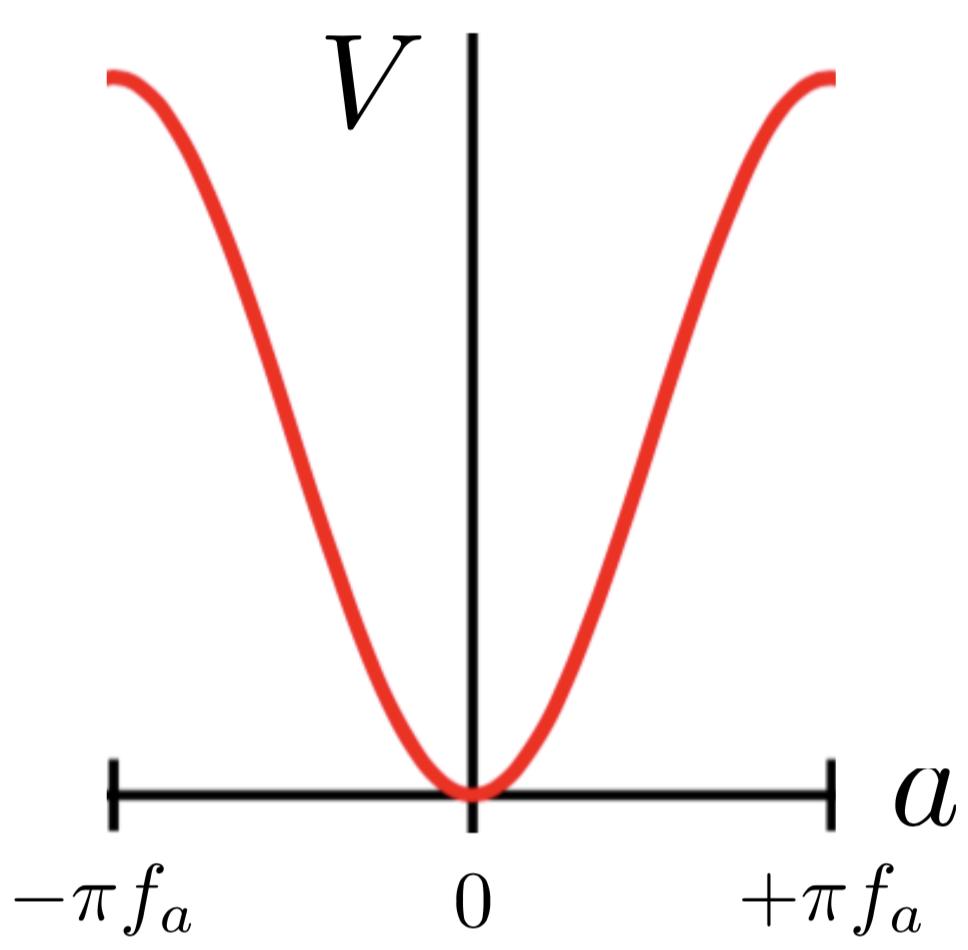}}
\caption[]{{\em Left:} Triangle loop diagram giving rise to the axion coupling to gauge fields. {\em Right:} Effective potential of the axion field.
\label{fig:triangle_potential}
}
\end{figure}
%
Specifically, their masses are given by $m_\rho =\sqrt{2\lambda_\sigma} v_{\rm PQ}$  and $m_{\mathcal Q} ={y} v_{\rm PQ}/{\sqrt{2}}$, respectively. Conversely, one particle, labeled as $a$, remains massless, serving as the Nambu-Goldstone boson arising from the breaking of the $U(1)_{\rm PQ}$ symmetry.
By integrating out the heavy particles $\rho$ and $\mathcal Q$, the dynamics of the massless field $a$ can be described by an effective Lagrangian: 
$$\mathcal L_{\rm eff} 
\supset \frac{1}{2} \partial^\mu a
\,\partial_\mu a + \frac{\alpha_s}{8\pi} \frac{a}{f_a}G_{\mu\nu}^a \tilde G^{a\,\mu\nu}.$$
The second term arises from the triangle loop diagram depicted in  Fig.~\ref{fig:triangle_potential} (left), from which one can read-off the factor of proportionality $\alpha_s/(8\pi f_a)$, where $f_a = v_{\rm PQ}/N_{\mathcal Q}$ is the axion's decay constant, with $N_{\mathcal Q}$ the number of 
PQ-charged quark flavors (one in the simplest scenario). 
Remarkably, this term assumes a form analogous to the $\theta$-term (\ref{eq:theta_term}) of QCD. Consequently, the 
$\overline\theta$-parameter can be rendered insignificant through a simple shift in the $a$ field: $a (x) +  \overline\theta\,f_a \to a (x)$. Subsequently, considering the mixing of the shifted $a$ field with the pion field, a non-trivial effective potential emerges for the former, featuring an absolute minimum at zero field value, as illustrated in Fig.~\ref{fig:triangle_potential} (right).
Put simply, the shifted  $a$ field exhibits a vacuum expectation value of zero, $\langle a \rangle =0$, effectively resolving the strong $CP$ puzzle~\cite{Peccei:1977hh}. The particle~\cite{Weinberg:1977ma,Wilczek:1977pj} associated with excitations of the shifted $a$ field has been termed the ``axion". Its mass is determined by the second derivative of the axion potential around the minimum, given by $m_a   = {\sqrt{V^{\prime\prime}(0)}} \simeq \frac{\sqrt{z}}{1+z}\,\frac{m_\pi \, f_\pi}{f_a} \approx 6\ {\rm \mu eV} \left( \frac{10^{12}\,{\rm GeV}}{f_a}\right)$, where $z=m_u/m_d\approx 1/2$, and $m_\pi$ and $f_\pi$  denote the mass and decay constant of the neutral pion, respectively.

\subsection{Electromagnetic coupling}

\begin{figure}[ht]
\centerline{\includegraphics[width=0.8\linewidth]{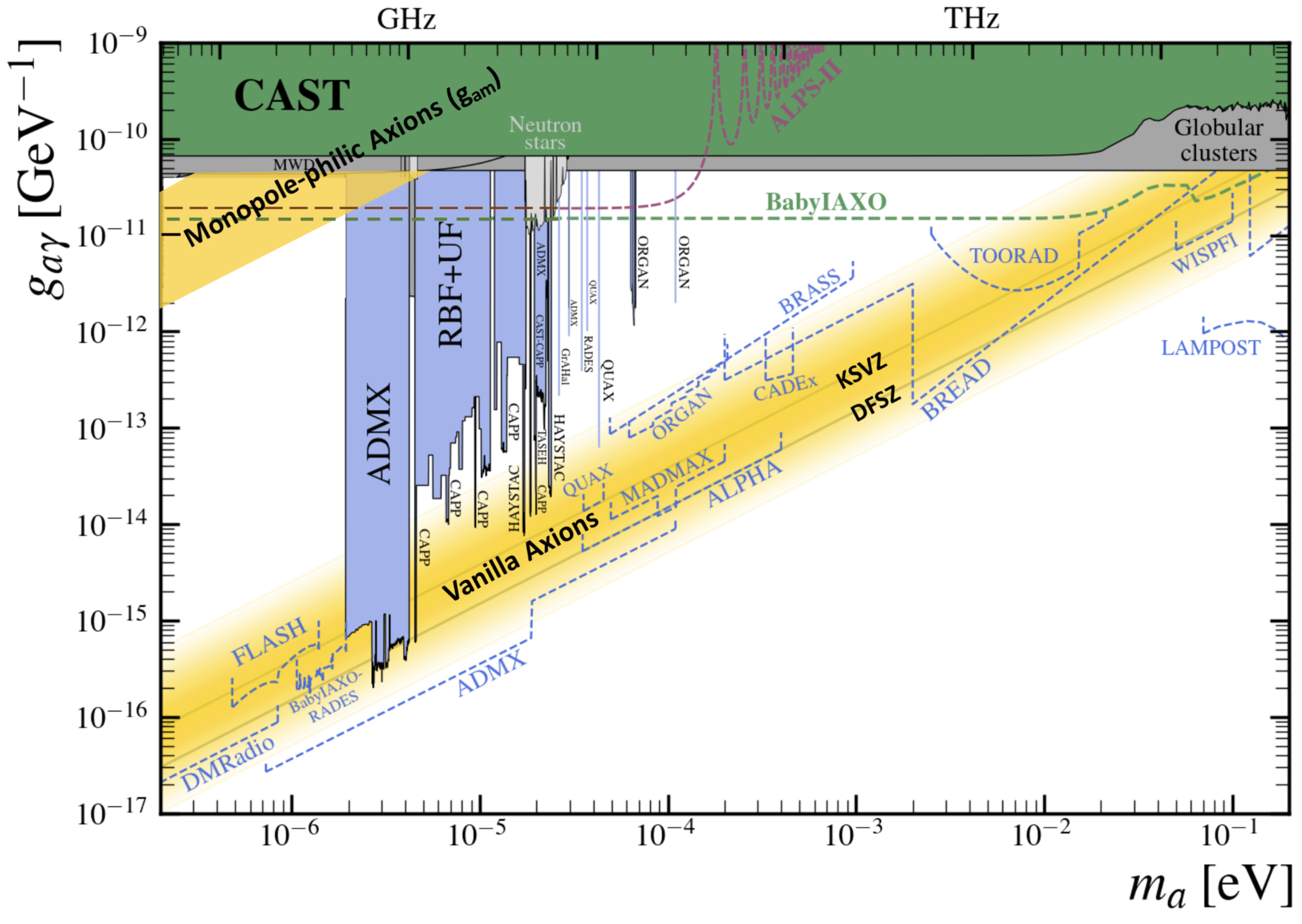}}
\caption[]{Electromagnetic coupling of the axion versus its mass (adapted from Ref.~\cite{AxionLimits}). Haloscope exclusion regions (filled blue) and projected sensitivities (dashed blue lines) assume the axion to be 100\% of the halo dark matter. 
\label{fig:electromagnetic_coupling}}
\end{figure}

Of paramount importance in both experimental and astrophysical~\cite{Caputo:2024oqc} realms is the axion's interaction with electromagnetic fields, described by the  term 
\begin{equation}
\mathcal L_{\rm eff} \supset \frac{1}{4} g_{a\gamma} \, a\, F_{\mu\nu} \tilde{F}^{\mu\nu}\equiv g_{a\gamma}\,a\,{\mathbf E}\cdot {\mathbf B}
\label{eq:electromagnetic_coupling}
\end{equation}
in its interaction Lagrangian, in terms of the (dual) electromagnetic field strength tensor $F_{\mu\nu}^b$ ($\tilde{F}^{b,\mu\nu}$) and the 
electric and magnetic fields, ${\mathbf E}$ and ${\mathbf B}$, respectively. 
 Here, the coupling parameter 
 $$g_{a\gamma} \simeq    \frac{\alpha}{2\pi f_\pi } \frac{m_a}{m_\pi}  \frac{1+z}{\sqrt{z}}
\left( {6\, q_{\mathcal Q}^{ 2}}{} - \frac{2}{3} \frac{4+z}{1+z} \right)$$ 
is proportional to the axion mass and the electromagnetic fine-structure constant,  
$\alpha = e^2/(4\pi)$. 
The first term within the brackets takes into account a possible electric charge $q_{\mathcal Q}$, in units of $e$, of the exotic quark $\mathcal Q$,
while the second term originates from the phenomenon of axion-pion mixing. 
In Fig.~\ref{fig:electromagnetic_coupling}, the line labeled `KSVZ' represents the predicted values of $g_{a\gamma}$ for $q_{\mathcal Q}=0$. When the electric charge $q_{\mathcal Q}$ is allowed to vary, it leads to a range of predictions depicted by the yellow `band', which is labeled as `Vanilla Axions' in Fig.~\ref{fig:electromagnetic_coupling}~\cite{DiLuzio:2016sbl}.

%
The alternate yellow `band' of predictions, denoted as `Monopole-philic Axions' in Fig.~\ref{fig:electromagnetic_coupling}, emerges under the condition where the KSVZ exotic quark carries a magnetic charge $g_{\mathcal Q}$, in units of the fundamental magnetic charge $g_0$, rather than an electric charge $q_{\mathcal Q}$, in units of the fundamental electric charge $e$,  In this scenario, the model not only resolves the strong CP puzzle but also addresses the challenge of charge quantization, as originally envisaged by Dirac, Schwinger, and Zwanziger~\cite{Dirac:1931kp,Schwinger:1966nj,Zwanziger:1970hk}.
The triangular loop diagram depicted in the left panel of Fig.~\ref{fig:triangle_potential} induces an electromagnetic coupling $g_{am}$, which can be approximated as 
$$g_{am} \simeq \frac{\alpha_m}{2\pi f_\pi } \frac{m_a}{m_\pi} \frac{1+z}{\sqrt{z}} 6 g_{\mathcal Q}^2,$$
where $\alpha_{m}$ denotes 
${g_0^2}/{(4\pi)}$~\cite{Sokolov:2021ydn,Sokolov:2021eaz,Sokolov:2022fvs,Sokolov:2023pos}.\footnote{It's worth mentioning that the phenomenological viability of this non-standard electromagnetic axion coupling has been recently challenged \cite{Heidenreich:2023pbi}, and our rebuttal can be found in an upcoming paper \cite{Sokolov:2024toapp}.}
The principle of charge quantization, represented by $e g_0=6\pi n$, where $n$ is an integer, leads to a parametric amplification of the coupling ratio, 
$$\frac{g_{am}}{g_{a\gamma}} = \frac{9}{4} \alpha^{-2} \left(\frac{g_{\mathcal Q}}{q_{\mathcal Q}}\right)^2 \sim 10^5,$$
as illustrated in Fig.~\ref{fig:electromagnetic_coupling}. This enhanced ratio is not feasible within another benchmark axion model proposed by Zhitnitsky~\cite{Zhitnitsky:1980tq} and Dine, Fischler, and Srednicki~\cite{Dine:1981rt}. In  this so-called  `DFSZ' model, the axion couplings to the SM gauge fields are generated by the SM quarks ($N_{\mathcal Q}=6$), none of which carry magnetic charges. Consequently, the electromagnetic coupling in this model falls within the lower boundary of the vanilla axion band depicted in Fig.~\ref{fig:electromagnetic_coupling}.

\section{Axion Experiments Not Relying on Axion Dark Matter}

\subsection{Light-shining-through-a-wall searches}

One powerful technique for producing and detecting axions in a controlled laboratory environment, independent of specific astrophysical models, is the ``Light-Shining-through-a-Wall" (LSW) technique~\cite{Anselm:1985obz,VanBibber:1987rq}. 
\begin{figure}[h]
\begin{minipage}{0.45\linewidth}
\centerline{\includegraphics[width=0.9\linewidth]{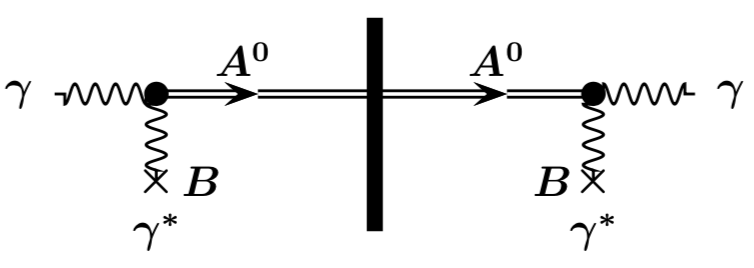}}
\end{minipage}
\hfill
\begin{minipage}{0.55\linewidth}
\centerline{\includegraphics[width=0.9\linewidth]{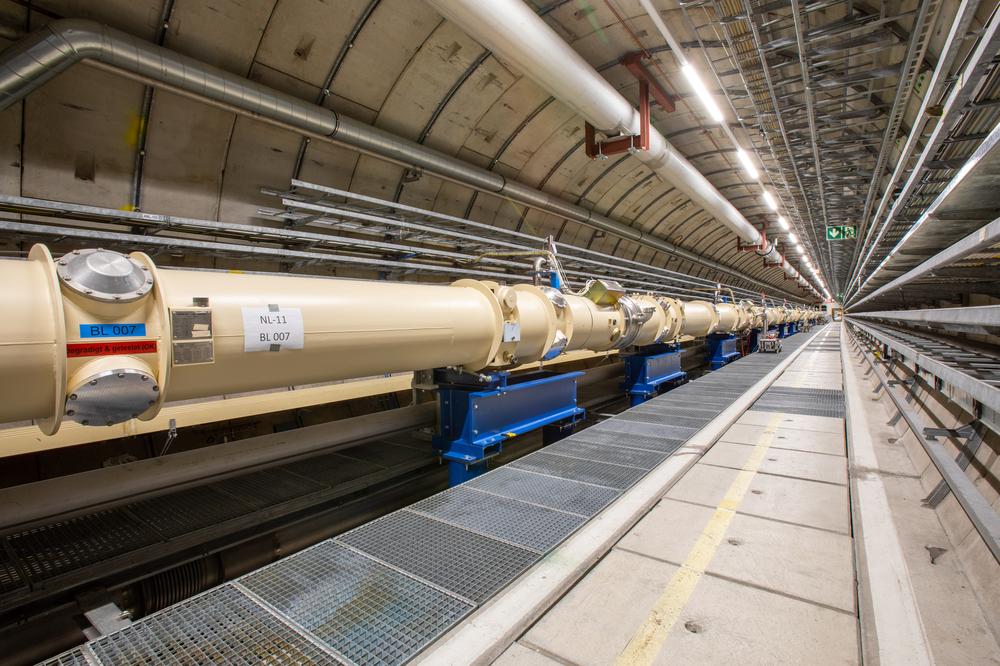}}
\end{minipage}
\caption[]{LSW concept~\cite{Ringwald:2003nsa} ({\em left}) and magnet string of the ALPS experiment in the HERA tunnel 
({\em right}).}
\label{fig:LSW_ALPS_II}
\end{figure}
%
This method exploits the phenomenon that photons traveling along a transverse magnetic field $B$ of length $L_B$ can undergo partial conversion into light axions, and vice versa, see Fig.~\ref{fig:LSW_ALPS_II} (left). For light axions, $m_a\ll (2\pi\omega/L_B)^{1/2}$, 
where $\omega$ is the photon energy, the probability of this conversion is approximately given by\footnote{Here, $g_{a\gamma}$ can be replaced by $g_{am}$ for the monopole-philic axion~\cite{Sokolov:2022fvs}.}  
$$P(\gamma \rightarrow a ) \simeq \frac{1}{4} \left( g_{a\gamma} B L_B\right)^2 \simeq P(a \rightarrow \gamma ).$$
By placing a light-tight barrier within the transverse magnetic field region, the presence of axions can be detected by observing photons emerging behind the barrier due to axion-photon conversion.
The ALPS II experiment~\cite{Bahre:2013ywa} at DESY in Hamburg employs several advancements to enhance sensitivity compared to previous LSW experiments~\cite{Redondo:2010dp}. It utilizes, as first proposed in Ref.~\cite{Ringwald:2003nsa},  two strings of recycled superconducting HERA dipole magnets -- 12 before and 12 after the wall -- within one of the straight sections of the HERA tunnel, see Fig.~\ref{fig:LSW_ALPS_II} (right),  opposed to just one magnet as in the preceding ALPS experiment~\cite{ALPS:2009des,Ehret:2010mh}. Additionally, ALPS II incorporates an optical cavity on the after-wall side to resonantly amplify the number of regenerated photons~\cite{Hoogeveen:1990vq,Fukuda:1996kwa,Sikivie:2007qm}, contrasting with the use of only one optical cavity before the wall as in ALPS. Data collection for ALPS II began in May 2023, with full sensitivity expected to be achieved by 2025. It promises to be approximately 1000 times more sensitive than previous LSW experiments such as ALPS and OSQAR~\cite{OSQAR:2015qdv}, thereby exploring previously uncharted parameter spaces, particularly for the monopole-philic axions, see Fig.~\ref{fig:electromagnetic_coupling}.

\subsection{Helioscope searches}

\begin{figure}[ht]
\begin{minipage}{0.49\linewidth}
\centerline{\includegraphics[width=\linewidth]{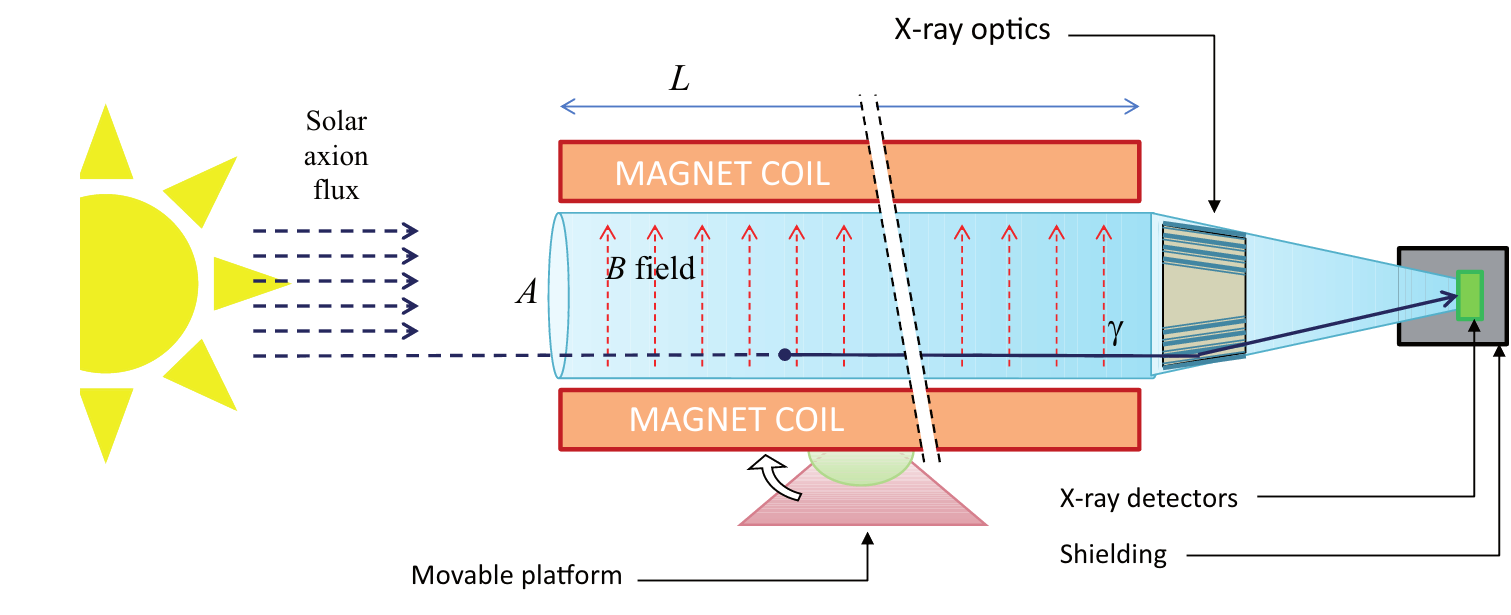}}
\end{minipage}
\hfill
\begin{minipage}{0.49\linewidth}
\centerline{\includegraphics[width=\linewidth]{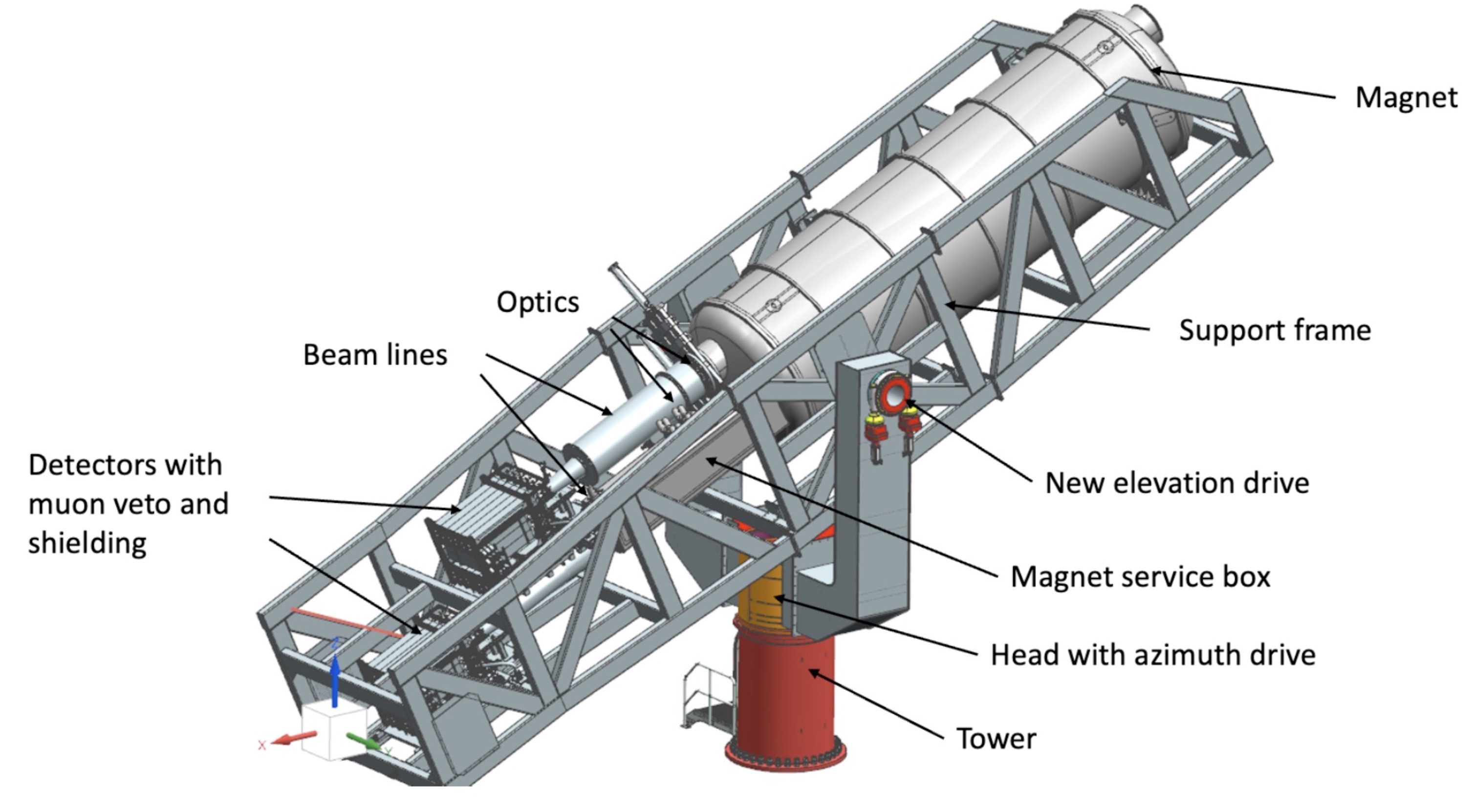}}
\end{minipage}
\caption[]{Axion helioscope concept~\cite{Armengaud:2014gea} ({\em  left}) and CAD overview of the full BabyIAXO assembly ({\em right}).}
\label{fig:BabyIAXO}
\end{figure}

Another avenue for axion exploration is through the detection of solar axions produced via the interaction of photons with the Coulomb field of nuclei in the solar plasma, achievable with an axion helioscope~\cite{Sikivie:1983ip}. This setup involves a long dipole magnet pointed towards the Sun, where solar axions are partially converted into photons and subsequently focused onto an X-ray detector, see Fig.~\ref{fig:BabyIAXO} (left).
%
%
CAST at CERN has set an upper limit~\cite{CAST:2017uph} on $g_{a\gamma}$ or $g_{am}$, which closely approaches the limit derived from observations of stars in globular clusters, see Fig.~\ref{fig:electromagnetic_coupling}. The next-generation helioscope, BabyIAXO~\cite{IAXO:2020wwp}, see Fig.~\ref{fig:BabyIAXO} (right), planned to be build in the HERA South Hall at DESY in Hamburg, is designed to surpass the sensitivity of CAST by a factor of approximately four in the same axion mass range, probing even the vanilla axion with masses exceeding 10 meV, see Fig.~\ref{fig:electromagnetic_coupling}. Data collection for BabyIAXO is anticipated to commence around 2029.

\section{Axion Dark Matter}

Just a few years following their original proposition, it was discovered that axions not only resolve the strong $CP$ puzzle but also emerge as promising candidates for cold dark matter~\cite{Preskill:1982cy,Abbott:1982af,Dine:1982ah}.
The prediction of the axion dark matter abundance critically hinges on the cosmic history, particularly on whether the PQ symmetry breaking occurs before or after the onset of the hot Big Bang phase at time $t_{\rm hot}$. In the former scenario, known as the `pre-inflationary PQ symmetry breaking scenario,' spatial gradients in the axion field $\theta_a (x)= a(x)/f_a$ can be disregarded, and its cosmic evolution can be fully described by a damped anharmonic oscillator. The damping arises from Hubble friction, while the anharmonicity stems from the specific form of the periodic axion potential, which, at temperatures surpassing the QCD cross-over, is represented as $V(\theta_a,T)=\chi (T) (1-\cos\theta_a)$, where $\chi (T)$ denotes the topological susceptibility of QCD, which can be obtained from high-temperature lattice QCD~\cite{Borsanyi:2016ksw}.
%
%
\begin{figure}[h]
\centerline{\includegraphics[width=0.7\linewidth]{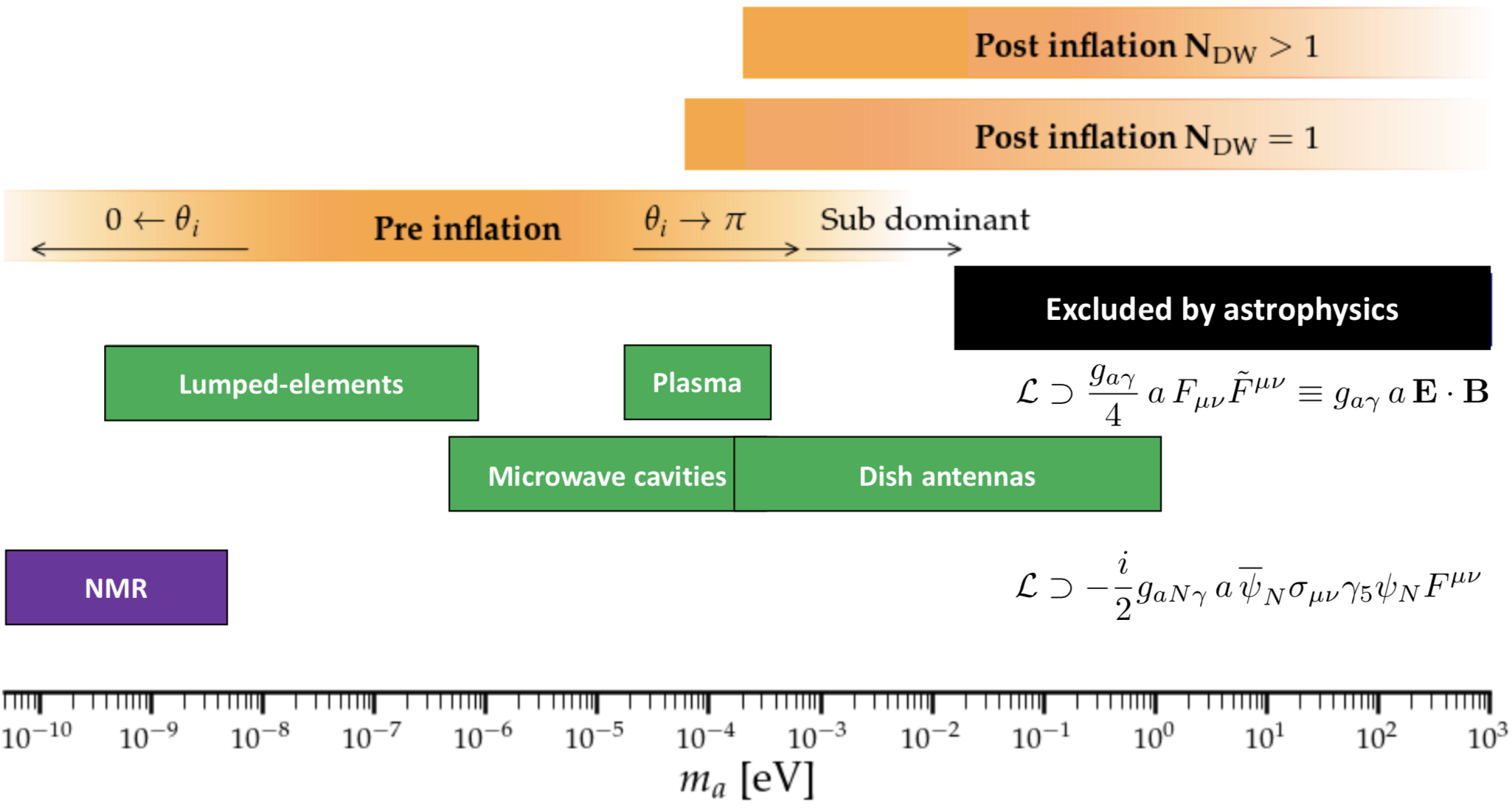}}
\caption[]{{\em Top three orange bands:} Theoretically predicted axion dark matter mass ranges.  {\em Middle black band:} Axion mass range excluded by astrophysics. {\em Green bands:} Axion dark matter ranges probed by various haloscopes. {\em Violet band:} Axion dark matter ranges probed by Nuclear Magnetic Resonance methods sensitive to NEDM oscillations or the axion dark matter wind.   Figure adapted from Ref.~\cite{Semertzidis:2021rxs}.}
\label{fig:axion_dark_matter_mass}
\end{figure}
%
The axion field remains fixed at its initial value, $\theta_{ai} \equiv a(t_{\rm hot})/f_a$, until the Hubble expansion rate falls below the axion mass $m_a(T)=\sqrt{\chi (T)}/f_a$. Subsequently, it oscillates around the $CP$-conserving minimum, representing a coherent condensate of cold dark matter.
The predicted abundance of axion dark matter, stemming from this mechanism called the `realignment mechanism,' is estimated as~\cite{Borsanyi:2016ksw} 
$$\Omega_ah^2
\approx 0.12\,\left({ 6~\mu{\rm eV}\over m_a}\right)^{1.165}\,
\theta_{ai}^2\,.$$
This prediction depends on both the axion mass and the initial value of the axion field. Hence, in the pre-inflationary scenario, the requirement for axion dark matter to avoid over-closure of the universe does not impose a lower bound on the axion mass, as illustrated in Fig.~\ref{fig:axion_dark_matter_mass}.
In contrast, in the `post-inflationary PQ symmetry breaking scenario,' where $\theta_{ai}$ may vary across different patches of the present universe, the realignment mechanism contributes an average amount~\cite{Borsanyi:2016ksw} 
$$\Delta
\Omega_ah^2\approx
0.12\,\left(\frac{30~\mu{\rm eV}}{ m_a}\right)^{1.165}.$$
Additionally, axion dark matter arises in this scenario also from the decay of cosmic strings and domain walls, yet their predicted contribution is subject to considerable uncertainties.
Consequently, the plausible range of axion masses capable of constituting 100\% of the observed dark matter in post-inflationary scenarios remains quite broad: $m_a \approx 26\ \mu{\rm eV} - 0.5\ {\rm meV}$ for axion models with short-lived domain walls, such as the KSVZ model ($N_{\rm DW}\equiv N_{\mathcal Q}=1$) \cite{Klaer:2017ond,Gorghetto:2020qws,Buschmann:2021sdq,Saikawa:2024bta}. For models featuring long-lived domain walls ($N_{\rm DW}>1$), like the DFSZ model with PQ symmetry arising from an accidental discrete symmetry, the predicted mass is notably higher \cite{Ringwald:2015dsf,Beyer:2022ywc}: $m_a \gtrsim {\rm meV}$.
Furthermore, substantial density fluctuations in the initial state of the axion field in the post-inflationary scenario lead to the formation of compact dark matter structures called ``miniclusters," contributing to heightened theoretical uncertainty in the local axion density for laboratory-based dark matter detection.

%
\begin{figure}[ht]
\centerline{\includegraphics[width=0.9\linewidth]{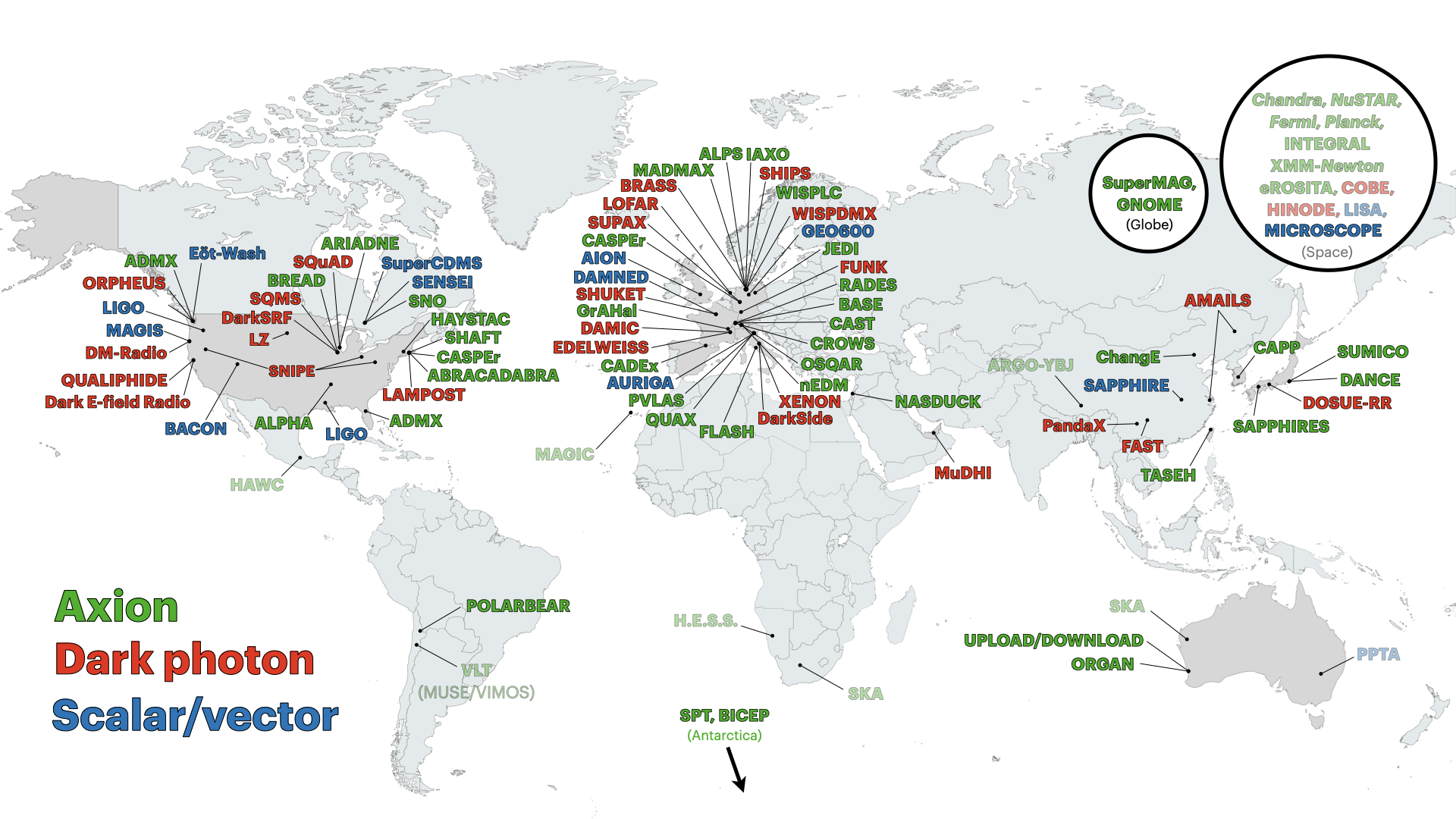}}
\caption[]{World map displaying current experiments searching for wavy dark matter~\cite{AxionLimits}.}
\label{fig:wavy_dark_matter_exp_world_wide}
\end{figure}
%

\section{Axion Dark Matter Experiments}

Axion dark matter experiments operate under the premise that the dark matter halo surrounding the Milky Way consists predominantly of axions. The velocity dispersion of these axions is determined by the galactic virial velocity, which in turn implies a macroscopic de Broglie wavelength, denoted as $\lambda_{\rm dB} = 2\pi/(m_a v_a) \simeq {\rm km}\, (\mu{\rm eV}/m_a)(10^{-3}\,c/v_a)$. Consequently, axions within the halo of dark matter behave akin to a spatially homogeneous and monochromatic classical oscillating field, characterized by the expression $a(t) \simeq \sqrt{2\rho_a} \cos (m_a t)/m_a$, thus falling into the category referred to as `wavy dark matter'.
In the context of presenting projected limits on the coupling of axions within the halo of dark matter, it is presumed that the axion energy density, denoted as $\rho_a$, is approximately equal to the energy density of dark matter in the halo, $\rho_{\rm DM}^{\rm halo} \approx 0.45\,{\rm GeV}\,{\rm cm}^{-3}$.

The current projections for the axion mass, potentially accounting for the observed dark matter content, encompass a wide range of values spanning multiple orders of magnitude, see Fig.~\ref{fig:axion_dark_matter_mass}. Consequently, investigating this extensive parameter space necessitates the utilization of various experimental techniques that may target different axion couplings.

An extensive array of axion experiments is currently underway worldwide, as depicted in Fig.~\ref{fig:wavy_dark_matter_exp_world_wide}. Given the sheer number of endeavors, it's evident that I can only focus on a subset of these experiments in the remainder of this brief overview.

\subsection{Haloscopes}

Numerous experiments investigating axion dark matter rely on the electromagnetic coupling (\ref{eq:electromagnetic_coupling})
and are thus referred to as `haloscopes'. These haloscopes employ various experimental techniques tailored to different mass ranges, as indicated by the shaded green regions in Fig.~\ref{fig:axion_dark_matter_mass}.

\subsubsection{Microwave cavities}

The original concept of the classic haloscope was introduced by Sikivie~\cite{Sikivie:1983ip}. In this setup, a microwave cavity is positioned within a magnetic field $B_0$, allowing the conversion of dark matter axions into photons.
If the axion's mass aligns with the resonance frequency of the cavity, denoted as $m_a = 2\pi \nu_{\rm res} \sim 4\,\mu{\rm eV} \left( \frac{\nu_{\rm res}}{\rm GHz}\right)$, the power output experiences amplification proportional to the cavity's quality factor $Q$ and volume $V$: $P_{\rm out} \sim g_{a\gamma}^2\, \rho_{\rm a}\, B_0^2\, Q\, V$. Given that the exact mass of the axion remains unknown, one has to scan the mass range of interest by tuning the cavity's resonance frequency. Currently, there exist numerous microwave cavity haloscopes in operation, among which ADMX~\cite{ADMX:2019uok} and CAPP~\cite{Yi:2022fmn} have achieved DFSZ sensitivity within certain mass ranges, as depicted in Fig.~\ref{fig:electromagnetic_coupling}.
Over the next decade, microwave cavity axion dark matter searches, such as ADMX~\cite{Stern:2016bbw}, BabyIAXO~\cite{Ahyoune:2023gfw}, and FLASH~\cite{Alesini:2023qed}, will delve deeper into the vanilla axion band within the range $0.5\,{\rm \mu eV} \lesssim m_a \lesssim 100\, {\rm \mu eV}$, as illustrated in Fig.\ref{fig:electromagnetic_coupling}. Regrettably, these searches remain insensitive to the non-standard coupling $g_{am}$ of the monopole-philic axions~\cite{Tobar:2022rko}.

\begin{figure}[h]
\begin{center}
\begin{minipage}{0.3\linewidth}
\centerline{\includegraphics[width=\linewidth]{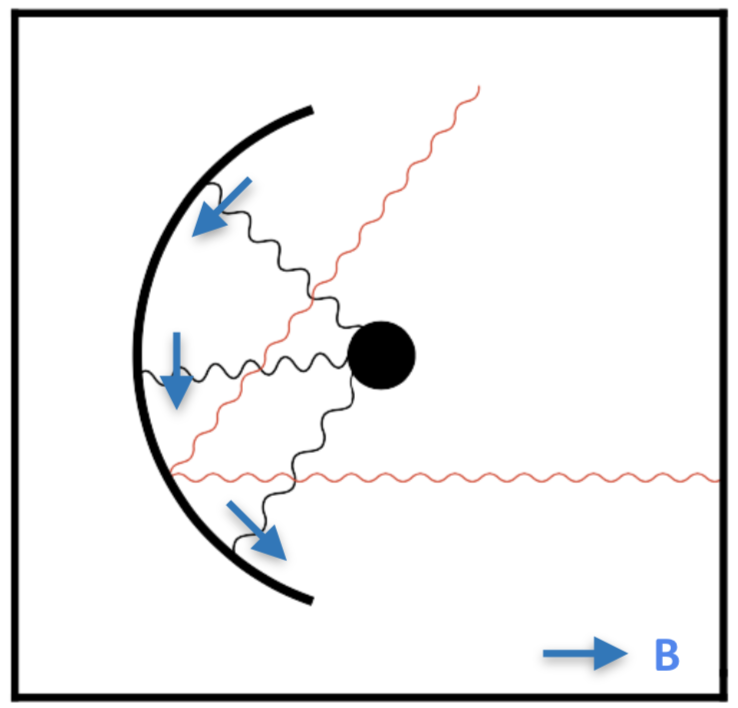}}
\end{minipage}
\hspace{12ex}
\begin{minipage}{0.45\linewidth}
\centerline{\includegraphics[width=\linewidth]{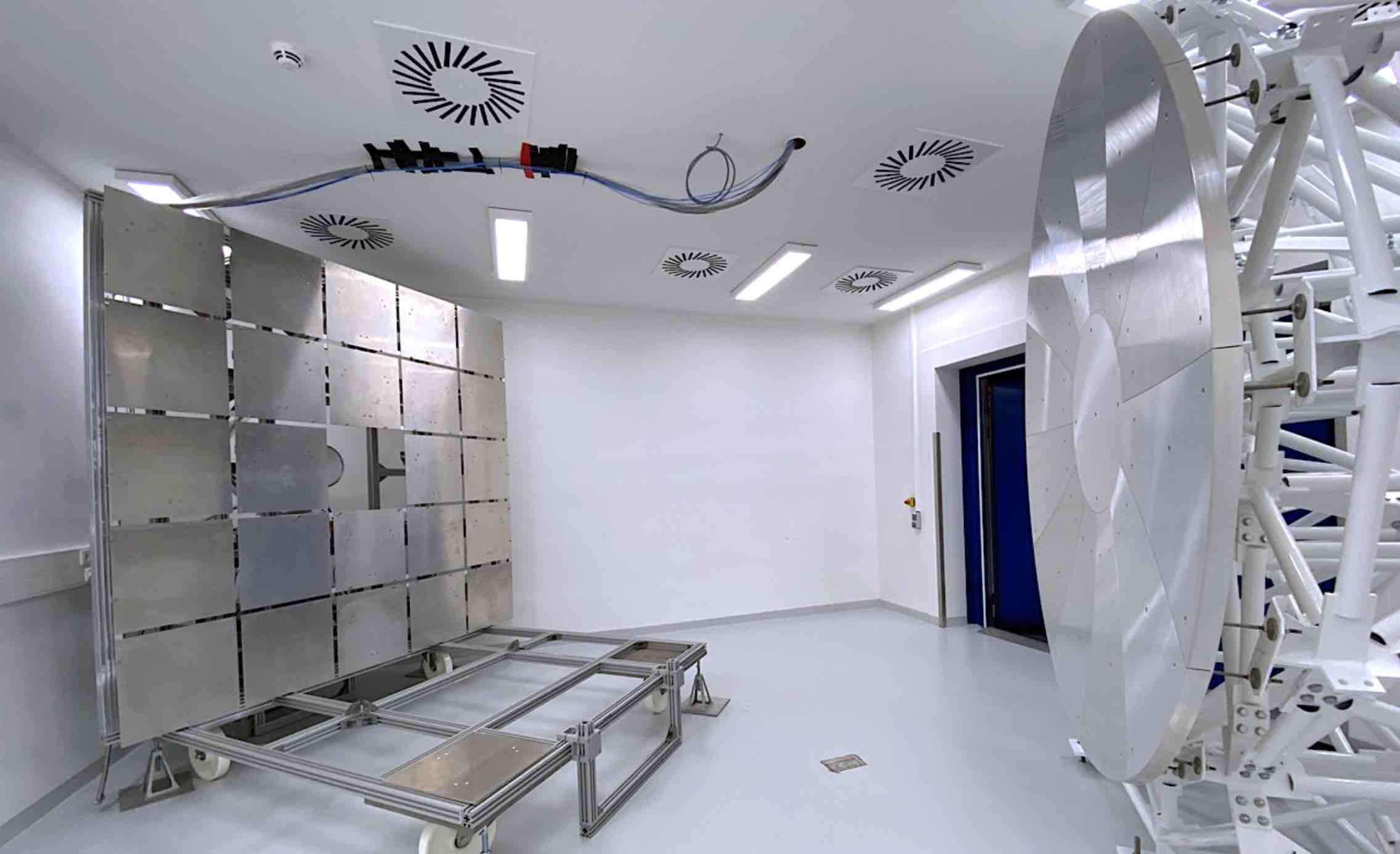}}
\end{minipage}
\end{center}
\caption[]{Dish antenna axion haloscope concept~\cite{Horns:2012jf}  ({\em  left}) and picture of BRASS ({\em right}).}
\label{fig:BRASS}
\end{figure}

\subsubsection{Dish antennas}

Exploring higher masses can be achieved through a broadband search approach facilitated by the dish antenna haloscope concept~\cite{Horns:2012jf}. This method capitalizes on the characteristic of oscillating axion dark matter, which, when subjected to a background magnetic field $\mathbf B$, generates an oscillating electric field component parallel to the magnetic field, denoted as ${\mathbf E}_a(t) = - g_{a\gamma}{\mathbf B} a(t)$. Consequently, when a metallic mirror is positioned in a magnetic field aligned parallel to its surface, it emits an almost monochromatic electromagnetic wave perpendicular to the mirror surface, as depicted in Fig.~\ref{fig:BRASS} (left), with a frequency $\nu = m_a/(2\pi)$ and a cycle-averaged power per unit area of ${{\mathcal P}_\gamma}/{\mathcal A}=|\mathbf{E}_a|^2/2$.
The BRASS collaboration at the University of Hamburg is presently establishing a pilot dish antenna haloscope utilizing a permanently magnetized conversion panel~\cite{Bajjali:2023uis}, see Fig.~\ref{fig:BRASS} (right). Meanwhile, the BREAD collaboration at Fermilab intends to utilize a cylindric parabolic conversion panel, enabling the employment of a much stronger solenoidal magnetic field~\cite{BREAD:2021tpx}. In their ultimate configuration, these experiments have the potential to explore the vanilla axion band region within the range $50\,{\rm \mu eV} \lesssim m_a \lesssim {\rm meV}$ (BRASS) or to delve deeply into it for $20\,{\rm meV} \lesssim m_a \lesssim 0.1\,{\rm eV}$ (BREAD), as illustrated in Fig.~\ref{fig:electromagnetic_coupling}.

\begin{figure}[h]
\vspace{-8ex}
\begin{minipage}{0.37\linewidth}
\centerline{\includegraphics[width=\linewidth]{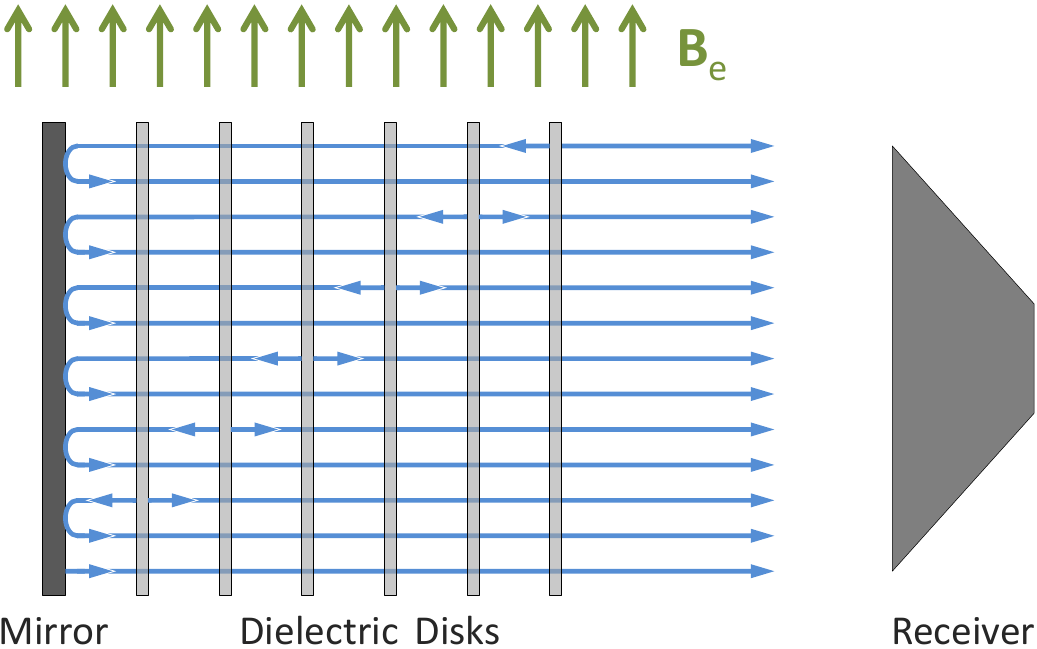}}
\end{minipage}
\hfill
\begin{minipage}{0.63\linewidth}
\centerline{\includegraphics[angle=270,width=\linewidth]{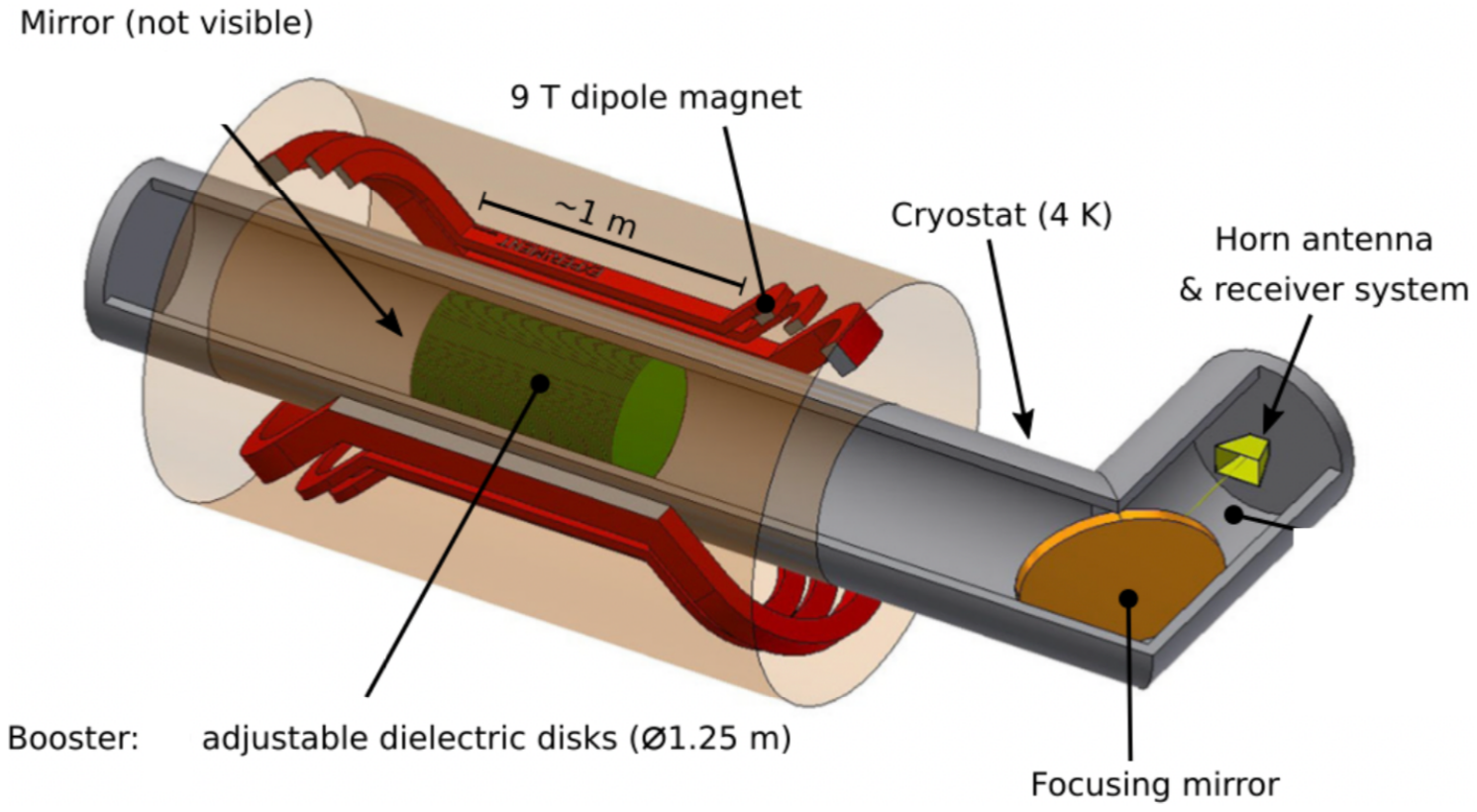}}
\end{minipage}
\vspace{-8ex}
\caption[]{Boosted dish antenna aka as 
dielectric haloscope concept~\cite{Caldwell:2016dcw} ({\em  left}) and sketch of MADMAX ({\em right}).}
\label{fig:madmax}
\end{figure}

An alternative method for detecting axion dark matter in the mass range above approximately $50\,{\rm \mu eV}$ is based on the dielectric haloscope concept~\cite{Caldwell:2016dcw}, essentially an enhanced version of the dish antenna haloscope. This setup comprises a mirror and a series of parallel, partially transparent dielectric disks placed in front of it, all positioned within a magnetic field parallel to their surfaces, with a receiver situated in a field-free region, see Fig.~\ref{fig:madmax} (left). Each disk functions as a flat dish antenna, emitting waves that are reflected by and transmitted through the other disks before exiting. With suitable separations between the disks, these waves coherently add up to amplify the emitted power. This approach enables scanning over a range of axion masses without requiring disks of varying thickness for each measurement.
Building upon this principle, the MADMAX collaboration aims to install an adjustable multiple-disk system (booster) with an effective area of approximately $\mathcal A \sim 1\,$m$^2$ within a dipole magnet of around $\sim 9$\,T at DESY in Hamburg~\cite{MADMAX:2019pub}, see Fig.~\ref{fig:madmax} (right). While data collection in Hamburg may commence around 2030, an earlier prototype magnet could facilitate axion dark matter searches in mass and frequency ranges that have largely remained unexplored.
With an anticipated power amplification factor of $\beta^2(\nu )\sim 10^4$ and equipped with a quantum-limited receiver, the complete MADMAX experiment is expected to survey the $(40-400)\,\mu$eV mass range with DFSZ sensitivity~\cite{Beurthey:2020yuq}, as depicted in Fig.~\ref{fig:electromagnetic_coupling}.

\subsubsection{Plasma haloscopes}

Another method suitable for exploring the mass range above $50\,{\rm \mu eV}$ is based on the plasma haloscope concept~\cite{Lawson:2019brd}. This technique capitalizes on the phenomenon that the oscillating axion dark matter field induces plasmon excitations in a magnetized plasma, expressed as 
${\bf E} = -g_{a\gamma}{\bf B}_{\rm e}a \left(1-\frac{\omega_p^2}{\omega_a^2-i\omega_a\Gamma}\right)^{-1}$. These excitations are resonantly amplified when the plasma frequency matches the axion mass, $\omega_p = \omega_a \approx m_a$, albeit constrained by losses ($\Gamma$). A plasma with a tunable plasma frequency in the GHz range can be achieved through a wire array with variable interwire spacing, often referred to as a ``wire metamaterial."
The ALPHA collaboration intends to construct a tunable cryogenic plasma haloscope along these lines at Yale. Currently, a pathfinder experiment is under development, expected to begin data collection in 2026. The full-scale ALPHA experiment is projected to delve deeply into the vanilla axion band within the $(40-400)\,\mu$eV mass range, as illustrated in Fig.~\ref{fig:electromagnetic_coupling}.

\begin{figure}[ht]
\begin{center}
\begin{minipage}{0.3\linewidth}
\centerline{\includegraphics[width=\linewidth]{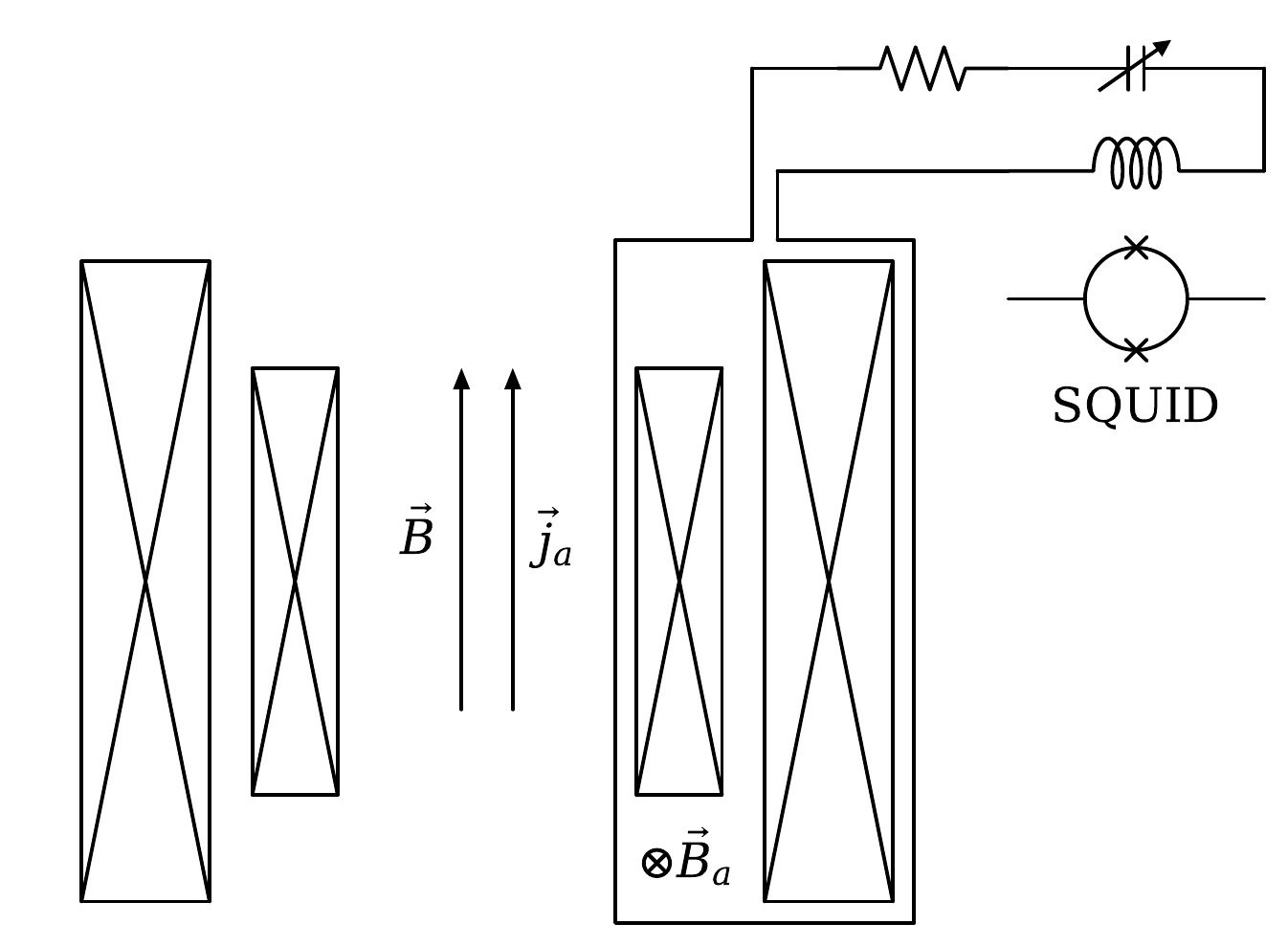}}
\end{minipage}
\hspace{12ex}
\begin{minipage}{0.18\linewidth}
\centerline{\includegraphics[width=\linewidth]{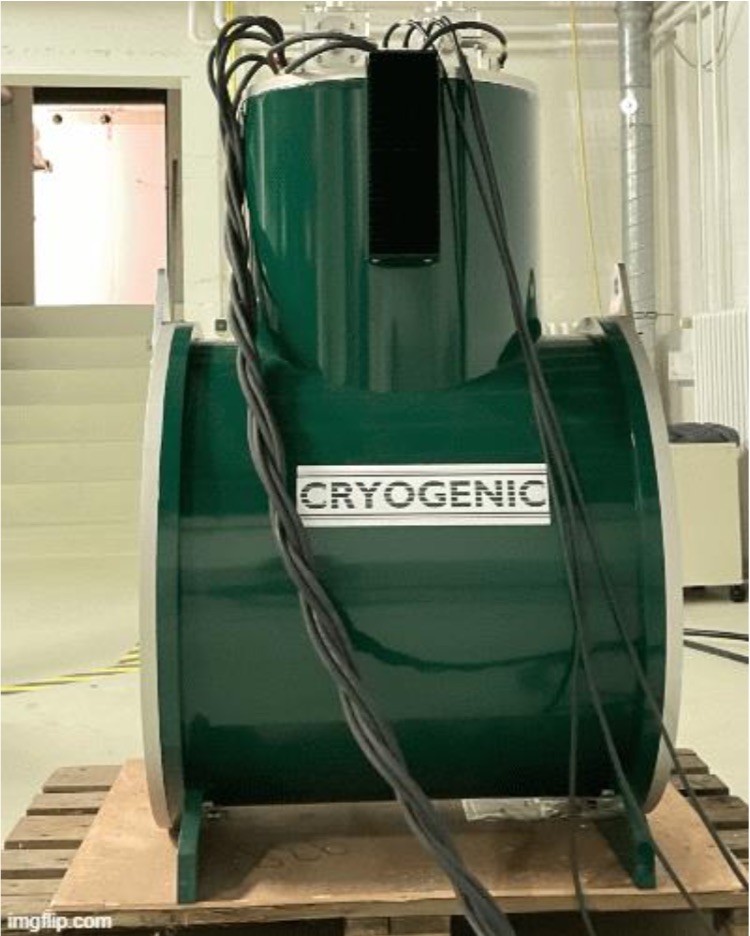}}
\end{minipage}
\end{center}
\caption[]{LC circuit haloscope concept~\cite{Zhang:2021bpa} ({\em  left}) and picture of the solenoidal magnet for WISPLC at the University of Hamburg  ({\em right}).}
\label{fig:WISPLC}
\end{figure}

\subsubsection{Lumped elements}

Around a decade ago, the concept of a lumped element axion haloscope emerged, offering sensitivity in the sub-micro-eV mass range~\cite{Sikivie:2013laa}. This approach is rooted in the fact that, under the influence of a magnetic field $\mathbf B$, the axion dark matter field triggers an oscillating effective displacement current, denoted as ${\mathbf j}_a = - g_{a\gamma} {\mathbf B} \dot a$, subsequently generating an oscillating magnetic field ${\mathbf B}_a$ such that ${\mathbf\nabla}\times {\mathbf B}_a = {\mathbf j}_a$. The induced field can be transformed into an alternating current within a pickup loop, resonantly amplified within a tunable LC circuit, and ultimately detected via a SQUID, as depicted in 
Fig.~\ref{fig:WISPLC} (left).
Pilot experiments have already been conducted along these lines, including ABRACADABRA~\cite{Ouellet:2018beu,Salemi:2021gck}, ADMX SLIC~\cite{Crisosto:2019fcj}, and SHAFT~\cite{Gramolin:2020ict}, while the next generation is currently commencing with projects like WISPLC~\cite{Zhang:2021bpa} (see Fig.~\ref{fig:WISPLC} (right)) and DMRadio~\cite{DMRadio:2022pkf}. The overarching objective of the DMRadio collaboration is to investigate axion dark matter within the neV to micro-eV range, aiming for DFSZ sensitivity, as depicted in 
Fig.~\ref{fig:electromagnetic_coupling}. Additionally, a variant of the lumped-element haloscope utilizing a high-voltage capacitor instead of a magnetic field may also explore the $g_{am}$ coupling of the monopole-philic axion~\cite{Li:2022oel,Tobar:2023rga}.

\subsection{NMR Experiments}

Lastly, we highlight an axion dark matter experiment that does not rely on the electromagnetic coupling but instead leverages the axion's least model-dependent coupling: its interaction with the nucleon electric dipole moment (NEDM) operator. This coupling is described by 
$$\mathcal{L}_{aN\gamma} = -\frac{i}{2} g_{aN\gamma}\, a\, \overline{\psi}N \sigma{\mu\nu} \gamma_5 \psi_N F^{\mu\nu},$$
where $g_{an\gamma}= - g_{ap\gamma} \approx 6\times 10^{-19} \left( \frac{m_a}{\rm neV}\right) \frac{1}{\rm GeV^2}$.
The oscillating axion field in the halo dark matter induces oscillations in NEDMs, expressed as $d_N (t) = g_{aN\gamma}\,\sqrt{2\rho_{a}} \cos (m_a\,t)/m_a$~\cite{Graham:2013gfa}. These induce precession of nuclear spins in a sample polarized with nucleon spins in the presence of an electric field. The resulting transverse magnetization can be detected by employing nuclear magnetic resonance (NMR) techniques, which are most sensitive at low oscillation frequencies corresponding to sub-neV axion masses, as indicated by the violet band in Fig.~\ref{fig:axion_dark_matter_mass}.
CASPEr-electric~\cite{Budker:2013hfa} in Boston is developing such an NMR axion search experiment~\cite{Aybas:2021nvn}. Its ultimate aim is to probe axion dark matter with neV masses, corresponding to a Peccei-Quinn (PQ) scale of approximately $10^{16}$\,GeV, as predicted in Grand Unified models (see, e.g., Refs.~\cite{Ernst:2018bib,DiLuzio:2018gqe,FileviezPerez:2019fku}).

\section{Conclusions}

In conclusion, there is a significant global effort underway in the search for axions, utilizing a variety of experimental techniques and exploring diverse couplings. The development of numerous novel experimental methods underscores the dynamic nature of this field, often stemming from close collaborations between theorists focused on phenomenological aspects and experimentalists driven by theoretical interest. As advancements continue to unfold, we encourage staying engaged and keeping abreast of the latest developments in this exciting area of research. Stay tuned for further updates!

\section*{Acknowledgments}
Special thanks to A.~Sokolov for valuable comments on the draft.
This work has been partially funded by the Deutsche Forschungsgemeinschaft (DFG, German Research Foundation) 
under Germany's Excellence Strategy -- EXC 2121 Quantum Universe -- 390833306  and under 
-- 491245950.
This article/publication is based upon work from COST Action COSMIC WISPers CA21106, supported by COST (European Cooperation in Science and Technology).

\end{document}